\documentclass[aps,prd,a4paper,twocolumn,showpacs,floatfix,nofootinbib,natbib]{revtex4-1}
\usepackage{graphicx}
\usepackage{subfigure}
\usepackage{amsmath}
\usepackage{amssymb}
\usepackage{sidecap}
\usepackage{color}
\usepackage{enumerate}
\usepackage{algorithm}
\usepackage{algpseudocode}
\usepackage{verbatim}
\usepackage{soul}
\usepackage[colorlinks,linkcolor=blue,citecolor=blue,urlcolor=blue ]{hyperref}
\def\ba{\begin{eqnarray}}
\def\ea{\end{eqnarray}}
\def\be{\begin{equation}}
\def\ee{\end{equation}}

\providecommand{\adsurl}[1]{\href{#1}{ADS}}

\newcommand{\snr}[1]{{\rm SNR}}

\begin{document}
\title{Testing Gravity with Gravitational Wave Source Counts}
\author{Erminia~Calabrese}\thanks{erminiac@astro.princeton.edu}
\address{Department of Astrophysical Sciences, Peyton Hall, Princeton University, 4 Ivy Lane, Princeton, NJ USA 08544}
\author{Nicholas~Battaglia}\affiliation{Department of Astrophysical Sciences, Peyton Hall, Princeton University, 4 Ivy Lane, Princeton, NJ USA 08544} 
\author{David~N.~Spergel}\affiliation{Department of Astrophysical Sciences, Peyton Hall, Princeton University, 4 Ivy Lane, Princeton, NJ USA 08544}

\begin{abstract}
We show that the gravitational wave source counts distribution can test how  gravitational radiation propagates on cosmological scales.  This test does not require obtaining redshifts for the sources. 
If the signal-to-noise ratio (SNR, $\rho$) from a gravitational wave source is proportional to the strain then it falls as $R^{-1}$, thus we expect the source counts to follow $dN/d\rho \propto \rho^{-4}$.  However, if gravitational waves decay as they propagate or propagate into other dimensions, then there can be deviations from this generic prediction.  We consider the possibility that
the strain falls as $R^{-\gamma}$, where  $\gamma=1$ recovers the expected predictions in a Euclidean uniformly-filled universe, and forecast the sensitivity of future observations to deviations from standard General Relativity. We first consider the case of few objects, 7 sources, with a signal-to-noise from 8 to 24, and impose a lower limit on $\gamma$, finding $\gamma>0.33$ at 95\% confidence level. The distribution of our simulated sample is very consistent with the distribution of the trigger events reported by Advanced LIGO. Future measurements will improve these constraints: with 100 events, we estimate that $\gamma$ can be measured with an uncertainty of $15\%$. We generalize the formalism to account for a range of chirp masses and the possibility that the signal falls as $\exp(-R/R_0)/R^\gamma$.

\end{abstract}

\maketitle

\section{Introduction}
Advanced LIGO's detection of gravitational waves (GW) has opened a new window into the universe.  This breakthrough detection will enable novel insights into the astrophysics of black holes, stellar evolution and fundamental physics.

Gravitational waves from supermassive black hole binary inspirals are powerful "standard sirens" with well defined luminosity distances~\cite{Holz2005}.  If we can identify astrophysical counterparts to the gravitational wave events, then the combination of redshift measurements and luminosity distances can be a powerful probe of cosmology and the nature of gravity~\cite{Schutz1986,Deffayet2007,ALigo2016GR}.

What if we can not identify the astronomical counterparts to gravitational wave events? This paper will emphasize that we can use the predicted source counts distribution as a probe of gravitational wave propagation. If the sources were at cosmological distances, then their statistical properties could be used as a probe of the geometry of the universe~\cite{Wang1997}.  However, even gravitational wave sources in the nearby universe can provide an important test of how gravitational waves propagate over large distances.  At $z << 1$,  we expect that the source counts should go as $dN/d\rho \sim \rho^{-4}$ where $\rho$ is the source strength or alternatively the signal-to-noise ratio (SNR) of the event.  However, if the gravitational wave signal decays or loses energy while travelling, perhaps due to propagation effects in  the bulk~\cite{Caldwell2001,Maartens2010}, or if the graviton can decay into composite particles~\cite{Hughes2001}, then this will alter the source counts distribution. Thus, even without identifying counterparts, we can use GW sources to constrain gravitational wave physics.

The paper is organized as follows. In Sec.~\ref{sec:GW} we summarize the gravitational wave definitions and observations useful for our work. We derive in Sec.~\ref{sec:counts} a model for the GW source counts that we will use in Sec.~\ref{sec:analysis} to constrain the propagation of the gravitational signal with simulations of future observations. We conclude and discuss our findings in Sec.~\ref{sec:conclude}.
 
\section{Gravitational Wave Detection}
\label{sec:GW}

Gravitational wave signals can be split into three broad groups: (i) transient signals with a duration between a millisecond and several hours (e.g., emitted by compact binary coalescences of two neutron stars (NS-NS), or a neutron star and a black hole (NS-BH), or two black holes (BH-BH)); (ii) long-duration signals continuously emitted (from e.g., spinning neutron stars); and (iii) stochastic backgrounds (arising from either cosmic inflation, cosmic strings, or from the superposition of unresolved sources). 

A signal is characterized by a dimensionless amplitude, strain or waveform, $h$, which will generate a fractional change in length across an interferometer.

The detection of a GW signal is obtained by comparing the data with some waveform templates (when available).
Searching for a small GW signal in a noisy environment will then depend on the strength of the signal as well as on how accurately the emission can be modelled, i.e., how accurately templates can be built. 
The gravitational waveform $h$ depends on the GW energy, $\Omega_{\rm GW}$, emitted by the source; its Fourier transform as function of frequency $f$ is~\cite{Moore2015}:
\be
|\tilde{h}(f)|^2= H_0^2\Omega_{\rm GW}(f)\frac{3}{8\pi^2}f^{-4} \,,
\ee
where $H_0$ is the Hubble constant.
$\Omega_{\rm GW}(f)$ is proportional to the energy spectrum, $dE(f)/df$, carried off from the source by the gravitational waves. 
High-significance detections will then be possible only for very loud sources or for emitters with well understood physics and accurate predictions for $dE(f)/df$. In the case of very uncertain theoretical waveform models (and weak signals), multi-messenger methods (e.g., measuring other associated emissions) will be the only way to pin down the GW signal. 

The theoretical prediction for $dE(f)/df$ are known with high accuracy (and depend on very few parameters) only for compact binary coalescences, making these objects the most promising sources of gravitational waves, with tens of events per year expected~\cite{Ligo2010,ALigo2016rate}. 
The GW emission will happen during three evolutionary phases of such binary systems: the two component objects will have an \emph{inspiral} regime (where the gravitational radiation is produced by loss of energy and angular momentum) that will continue until a stable orbit is reached; the objects will then start a \emph{merging} phase leading to a plunge into a single component (single black hole or hypermassive neutron star); at last the final object will settle into a Kerr stationary quiescent state after undergoing a \emph{ringdown} phase. The chirp mass (a non-linear combination of the masses of the two initial objects) of the system is the main parameter determining amplitude and frequency evolution of the gravitational wave signal.

The main method used to detect GW emission from binary systems is to built an optimally match-filtered signal-to-noise ratio ({\rm SNR}) by integrating the data against theoretical waveforms divided by the spectral noise density $S_n(f)$ of the detectors. The average SNR over an ensemble of noise realization at fixed incident waveform is given by~\cite{Wainstein1962,Cutler94, Flanagan97}: 
\be
{\rm SNR}^2 = 4 \int_0^\infty \frac{|\tilde{h}(f)|^2}{S_n(f)} df \,.
\label{eq:sn}
\ee

For its current analysis, the LIGO team is requiring that 
detections must have {\rm SNR} above a certain cut defined by a small probability of false events, in both of its  detectors. A minimum value, ${\rm SNR}_{\rm min}$, will set a threshold for trigger events. As the gravitational wave experiments grow in number and sensitivity, this threshold will likely evolve. 

The signal-to-noise from a source will however depend on the distance to the source and its chirp mass. The detector noise will in fact determine the strength of the weakest signals that can be detected, and thus the distance to which a given type of source can be seen. 
In the case of the advanced detectors currently at work, NS-NS, NS-BH, BH-BH systems, with typical black hole masses up to a few tens of solar masses will be visible out to hundreds of Mpc~\cite{Miller2015}.

Assuming that the source population is local, the GW signal-to-noise ratio scales inversely with the luminosity distance and follows a universal distribution function of the detection threshold~\cite{Schutz2011,Chen2014}. If the source distribution covers a wide range of redshift, then both cosmology and source evolution will alter the events distribution.

\subsection{Observations from Advanced LIGO}

Advanced LIGO \cite{ALigo2015} is one of the ground-based experiments contributing to a world-wide network of detectors (including LIGO \cite{ALigo2015}, Virgo \cite{Virgo2015} and GEO600 \cite{Geo2010} currently operating, and with KAGRA \cite{Kagra2012} and Ligo-India \cite{India2011} joining in the next few years) designed to capture the gravitational wave signal. To extend the success of the initial and Enhanced LIGO science runs (see e.g., \cite{Abbott2009,Abadie2010a,Abadie2010,Abadie2011,Abadie20120,Abadie2012a,Abadie2012,Aasi2013,Aasi1,Aasi2}), Advanced LIGO will gradually lower the sensitivity by approximately a factor of ten relative to LIGO, corresponding to a detection volume a factor of about a thousand bigger. Moreover, the improved sensitivity will open a new window for detection at low frequencies, extending the low end of the band from 40 Hz to 10 Hz. 

Advanced LIGO had a first four months observing run from September 2015.  The improvement in sensitivity successfully resulted in the high-significance detection (${\rm SNR}\sim 24$) of a merging BH-BH system, GW150914, followed during the full evolution from the inspiral to the ringdown from both the Hanford and Livingston detectors~\cite{ALigo2016}. Other 6 trigger events were also observed at lower significance ($8<{\rm SNR}<10$), with one other event marginally detected above background~\cite{ALigo2016sources}. The detections are reported from matched-filter analyses for systems with low false rate probability and above a threshold for triggers of ${\rm SNR}_{\rm min}$$=8$. 

\section{Source Counts as a Test of Gravitational Wave Physics}
\label{sec:counts}

General Relativity (GR) has been extremely successful in describing an enormous number of gravity processes over a wide range of scales and many cosmic epochs (see \cite{Baker2015} for a review). Gravitational waves contribute to testing gravity by accessing the most extreme gravitational regimes as well as non-luminous mysterious objects, with the additional advantage of travelling from the source to the observer without interference or attenuation (in standard models of gravity)~\cite{ALigo2016GR}.
We derive here a parametrization offering a new method to test GW theory by probing the propagation of the gravitational wave signals in the local universe. 

\subsection{Source counts scaling predictions}

In the standard theory of gravity, in a uniformly-filled universe, the signal-to-noise associated to a low-redshift GW source ($z<<1$) at a distance R scales as:
\be
\rho \propto 1/R \,,
\ee
so that:
\ba
R(\rho)&\propto& \frac{1}{\rho}\,, \\
dR&\propto&-\frac{1}{\rho^2} d\rho \,.
\ea

The radial distribution of the sources should be Euclidean and follow:
\begin{equation}
\frac{dN}{dR} \propto R^2 \,.
\label{eq:euc}
\ee

We can now write the expected source counts using the chain rule:
\ba
\frac{dN}{d\rho} &=& \frac{dN}{dR}\frac{dR}{d\rho}   \nonumber \\
& = & A \rho^{-4} \,,
\ea
where $A$ is a constant. $A$ encodes the dependence on the source population that we have integrated out here because it does not affect the slope of the distribution.

In the case of departures from the standard propagation of the signal, let's assume that the signal-to-noise scales with a $\gamma$ power:
\ba
\rho &\propto& 1/R^\gamma \,, \\
R(\rho)&\propto&\frac{1}{\rho^{1/\gamma}} \,, \\
dR&\propto&-\frac{1}{\gamma} \rho^{-1/\gamma-1} d\rho\,.
\ea

The differential number counts will now be:
\be
\frac{dN}{d\rho}=A(\gamma) \rho^{-3/\gamma-1} \,,
\label{eq:modcounts}
\ee
where the normalization constant is now degenerate with the $\gamma$ slope of the function.

Finally, we choose to renormalize the counts and work with:
\ba
\frac{dN}{d\rho}|_{\rm norm}&=&C \frac{dN}{d\rho}/\frac{dN}{d\rho_{\rm min}} \nonumber \\
&=&C \Big ( \frac{\rho}{\rho_{\rm min}}\Big )^{-3/\gamma-1} \,,
\label{eq:counts}
\ea
where $C$ is a constant depending on the number of sources observed and their distribution ($C\sim3/\gamma \times N_{\rm objects}$). $\rho_{\rm min}$ is the minimum signal-to-noise ratio with which the data will be reported. We will refer to $dN/d\rho|_{\rm norm}$ as $dN/dx$ with $x=\rho/\rho_{\rm min}$. 

With this method we can pin down possible departures from the standard GW theory if in the modified theory of gravity the gravitational wave signal gets modified while travelling and the power law index deviates from the prediction of $\gamma=1$. 

This derivation breaks down if the source sample extends to intermediate-to-high redshifts: in that case evolution in the source distribution, and in the cosmology, will modify how the SNR scales with distance. Departures from $\gamma=1$ could indicate either non standard propagation of gravity or evidence for redshift evolution. With current knowledge we cannot derive accurate predictions for the source evolution and hence in this paper we will restrict the source sample to objects at $z<<1$ so that the assumption of uniform distribution is a good approximation. However, even in the case of future observations extending to higher redshift, one might choose only low chirp mass events to select local sources and work with this simplified derivation.

\subsection{Generalized derivation including dependence on chirp mass and signal cut-off}
The source distribution might show deviations from the standard power law predictions also if a signal cut-off is present.
We now generalize the derivation of the previous section to include the dependence on the chirp mass of the binary system and a possible exponential cut-off in signal from a source.

As we noted above, for sources at low redshift, their radial distribution should be Euclidean (from Eq.~\ref{eq:euc}) and their distribution of chirp masses should be independent of $R$:
\begin{equation}
\frac{dN}{dM_c} \propto g(M_c) \,,
\end{equation}
where $g(M_c)$ is some unknown function that describes the chirp mass.  The signal also depends on
other properties of the inspiral - we could generalize this derivation by replacing $M_c$ with a vector
that contains all of the physical parameters that determine the signal-to-noise of the event.
For each chirp mass $M_c$, we can define a maximum radius out to which a signal can be detected over the
threshold, $R_{\rm max}(M_c)$, so that the {\rm SNR} can be written as:
\begin{equation}
\rho = \frac{R_{\rm max}(M_c)}{R} \,.
\label{eq:snmax}
\end{equation}

Following the formalism introduced before, in a non-standard GW theory, the above equation becomes:
\begin{equation}
\rho = \frac{R_{\rm max}(M_c)}{R^\gamma} \,.
\end{equation}

The source counts can now be written as:
\begin{eqnarray}
\frac{d^2N}{dM_c d\rho} &=& \frac{dN}{dM_c}\frac{dN}{dR}\frac{dR}{d\rho} \nonumber \\
&=& \frac{\alpha g(M_c) R_{\rm max}(M_c)^{3/\gamma}}{\rho^{3/\gamma+1}} \,,
\end{eqnarray}
where $\alpha$ is a constant.

Integrating over $M_c$ we get again Eq.~\ref{eq:modcounts}:
\begin{equation}
\frac{dN}{d\rho} = B \rho^{-3/\gamma-1} \,,
\end{equation}
where B is a constant. Non-standard gravity corrections to $R_{\rm max}$ are absorbed by the normalization constant and do not impact the slope.

We now generalize further this derivation to account for an exponential cut-off in the signal. We consider an exponential attenuation in the signal with a characteristic decaying length-scale $R_0$ and derive the source distribution:
\begin{equation}
\rho = \frac{R_{\rm max}(M_c)}{R}\exp(-R/R_0) \,.
\end{equation}

This now yields an equation for $R(\rho,M_c)$.
Noting that:
\begin{equation}
\frac{dR}{d\rho} = -\frac{1}{\rho} \frac{R R_0}{R+R_0} \,,
\end{equation}
the source counts distribution can be written as:
\begin{equation}
\frac{d^2N}{dM_c d\rho} = \frac{\alpha g(M_c)}{\rho} \frac{R(M_c,\rho)^3 R_0}{R(M_c,\rho) + R_0} \,.
\end{equation}

We can now introduce again non-standard GW propagation and write:
\begin{equation}
\rho = \frac{R_{\rm max}(M_c)}{R^\gamma}\exp(-R/R_0) \,,
\end{equation}
with:
\begin{equation}
\frac{dR}{d\rho} = -\frac{1}{\rho} \frac{R R_0}{R+\gamma R_0} \,,
\end{equation}
and the source counts distribution becomes:
\begin{equation}
\frac{d^2N}{dM_c d\rho} = \frac{\alpha g(M_c)}{\rho} \frac{R(M_c,\rho,\gamma)^3 R_0}{R(M_c,\rho,\gamma) + \gamma R_0} \,.
\end{equation}

This extended derivation will allow us to look for both non-standard GW propagation and a possible signal cut-off affecting the source distribution. However, the integral over $M_c$ now can not be done analytically so we would need to evaluate
the likelihood for each event as a function of $M_c$ and SNR. This will be possible only once Advanced LIGO has provided the maximum detection distance as a function of black hole binary parameters. Non-standard gravity corrections should be then consistently propagated to the derivation of $R_{\rm max}$.

\section{Analysis}
\label{sec:analysis}
We simulate observations of differential number counts from the cumulative distribution of sources detected at a given SNR above a threshold ${\rm SNR}_{\rm min}$. We choose ${\rm SNR}_{\rm min}$$=8$ based on the minimum SNR for triggers in Advanced LIGO searches. The minimum SNR value does not affect our results and when real data of many detections will be available we will work with the reported SNR and threshold of the detections.
We estimate the power law scaling of the source distribution by comparing the simulated source counts with the theory prediction of Eq.~\ref{eq:counts}. 

\subsection{Likelihood and Sampling}
Source counts follow Poisson statistics and are described by a probability distribution:
\be
P({d_i}|{\lambda_i})=e^{-\lambda_i}\frac{\lambda_i^{d_i}}{{d_i}!} \,,
\ee
where, after dividing the data in $N_b$ bins, $d_i$ is the number of objects in a bin $i$ and $\lambda_i$ is the mean value of $dN/dx$ in the bin.

The log-likelihood function describing our sample is then \cite{Cash79}:
\ba
ln\mathcal{L}&=&ln\prod_{i=1}^{N_b}  P({d_i}|{\lambda_i})\\
&=&\sum_{i=1}^{N_b}(d_i ln(\lambda_i) - \lambda_i) \,.
\ea

We ignore corrections to the likelihood due to cosmic variance, considering its contribution negligible within the volume covered by our sample.

By tuning the size of the bins such that each bin contains only one object, we calculate the $ln\mathcal{L}$ function per single object. This avoids the compression of information into bins which is sub-optimal for parameter estimation. 

We use \texttt{emcee} \cite{FM2013} to sample the $ln\mathcal{L}$ function and to map out the posterior distribution of the parameters of our model, $C$ and $\gamma$. We then extract from the chain marginalized distributions for each parameter. In particular, we are only interested in constraining the model by measuring the slope of the distribution (and hence the gravity scaling) and $C$ is not an informative parameter. The exact value of $C$ will vary with the sample and we simply marginalize over it. We impose a flat prior on both parameters with $C$ in the range [0,500] and $\gamma$ varying in [0,5]. The lower bounds of our ranges are motivated by physical assumptions - we do not allow for negative counts and for inverted gravity - while the upper bounds are values that allow to sample the parameters space without biasing the results, i.e., covering all of the correlated region. We have however checked that our results are robust and are unchanged if we leave more freedom in the MCMC excursion.  

\subsection{Results}
\begin{figure}[t!]
\centering
\includegraphics[width=\columnwidth]{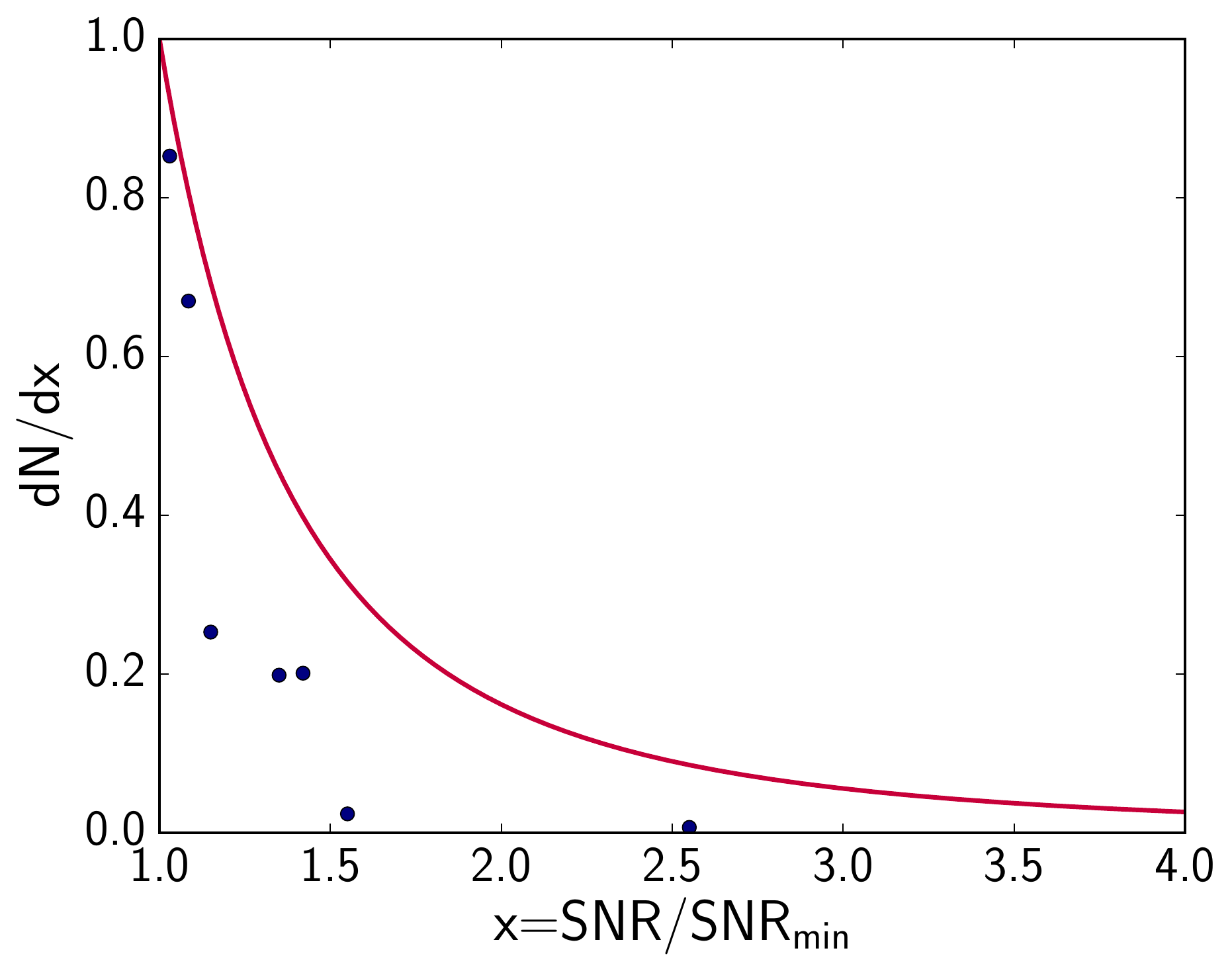}
\caption{Differential source counts as function of normalized SNR. The blue datapoints are 7 simulated observations with $8<{\rm SNR}<24$. The red curve is the model best-fitting the data. We normalize both the data and the model to have $dN/dx=1$ at $x=1$ and 0 at $x\to\infty$.
\label{fig:databf}}
\end{figure}

\begin{figure}[t!]
\centering
\includegraphics[width=\columnwidth]{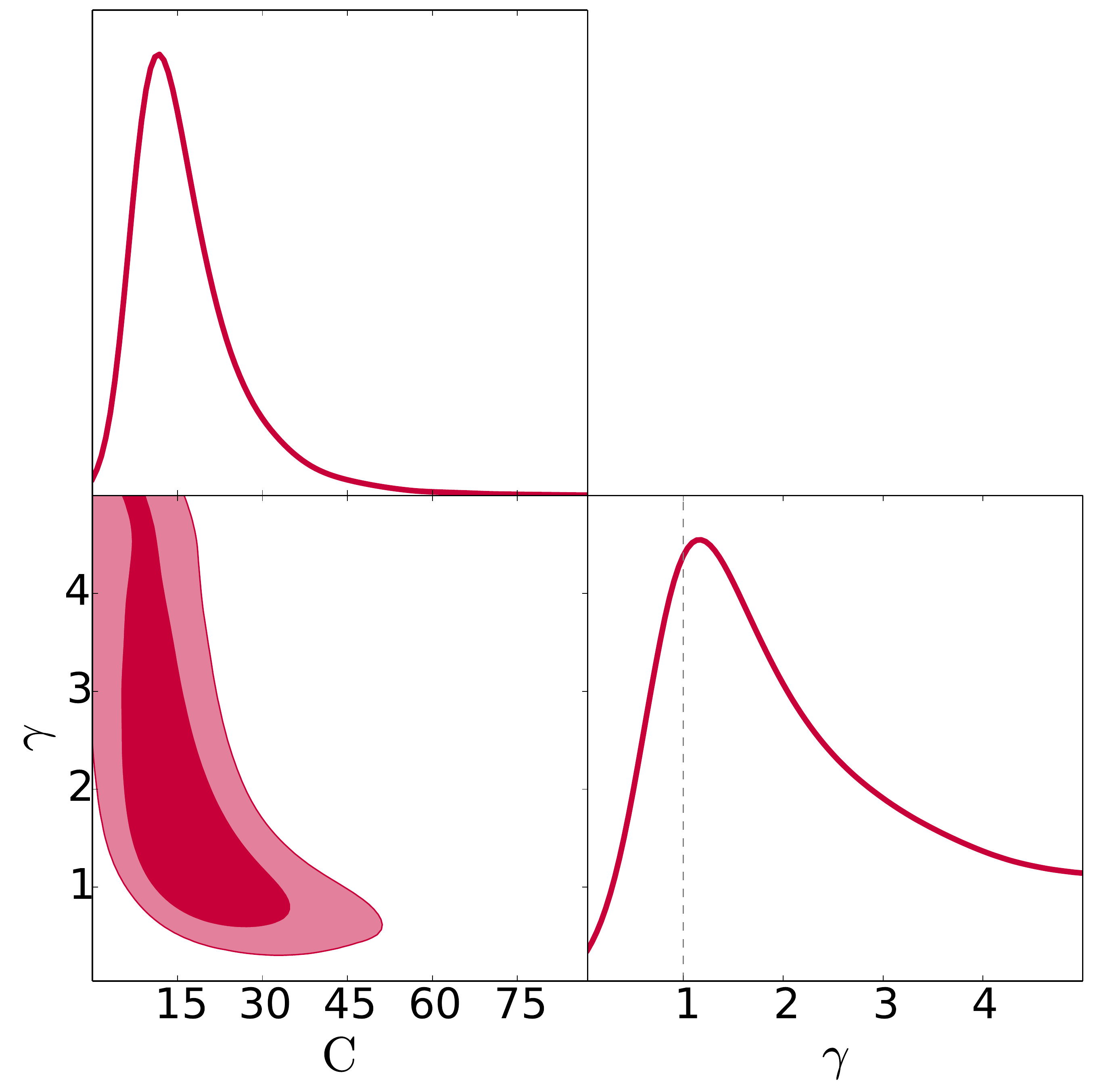}
\caption{1-dimensional posterior distributions and 2-dimensional contours showing the $68\%$ and $95\%$ confidence levels for the amplitude $C$ and the slope parameter $\gamma$ from 7 sources. The vertical dashed line shows the expected prediction for standard GW propagation.
\label{fig:tridata}}
\end{figure}

\subsubsection{Few detections}
We first consider the case of only few objects detected. We use the GW event with ${\rm SNR} \sim24$ reported by Advanced LIGO to predict the number of other events at lower significance expected in the standard scenario. Given Eq.~\ref{eq:sn}, we expect:
\be
<\rho_{\rm max}> = A(\rho_{\rm max})^{-4}\exp\Big(-\frac{1}{3}A(\rho_{\rm max})^{-3}\Big) \,.
\ee 

We estimate the rate $N_0$ by setting the mean of the distribution $<\rho_{\rm max}>=\Gamma(2/3)N_0^{1/3}/3^{1/3}$. Thus we expect $dN/d\rho\sim1.2 \rho_{\rm max}^3/\rho^4$ and $\sim10$ events above a ${\rm SNR}_{\rm min}$$=8$ threshold. We then choose to run our pipeline on a first case of only 7 objects observed with $8<{\rm SNR}<24$. This simulation is very consistent with the distribution of the trigger events observed by Advanced LIGO. 

The observed counts with the model best-fitting the data are reported in Fig.~\ref{fig:databf}. Here, we choose to normalize both the data and the model to have $dN/dx=1$ at $x=1$ and 0 at $x\to\infty$. The 1-dimensional posterior distributions and 2-dimensional contours showing the $68\%$ and $95\%$ confidence levels for the parameters of our model are reported in Fig.~\ref{fig:tridata}. The resulting distributions show that the low number of sources prevents a definitive measurement of both the slope and the amplitude, a higher amplitude is possible at the expenses of a tilt in slope of the model. With only 7 objects one can, however, already impose a lower bound on $\gamma$ finding a marginalized value of $\gamma>0.33$ at 95\% confidence level, in agreement with the expectation of $\gamma=1$.

\subsubsection{Many detections over a wide SNR range}
To forecast the improvement in the constraints achievable with the detections of many more sources, as we expect from future data from LIGO and VIRGO, we simulate 100 observations for standard GW propagation ($\gamma=1$) spanning a wider SNR range, $8<{\rm SNR}<\infty$ ($1<x<\infty$), and re-run our pipeline. With a factor of $\sim10$ more in the number of detected sources and with the data extending to larger SNR (then probing the slope of the curve over a wider range) the standard scaling is recovered and constrained with $\sigma(\gamma)\sim0.15$ (see Fig.~\ref{fig:trisim}). In this case the data are also able to break the degeneracy between the amplitude and the slope and measure clearly both parameters at the same time. 

\begin{figure}[t!]
\centering
\includegraphics[width=\columnwidth]{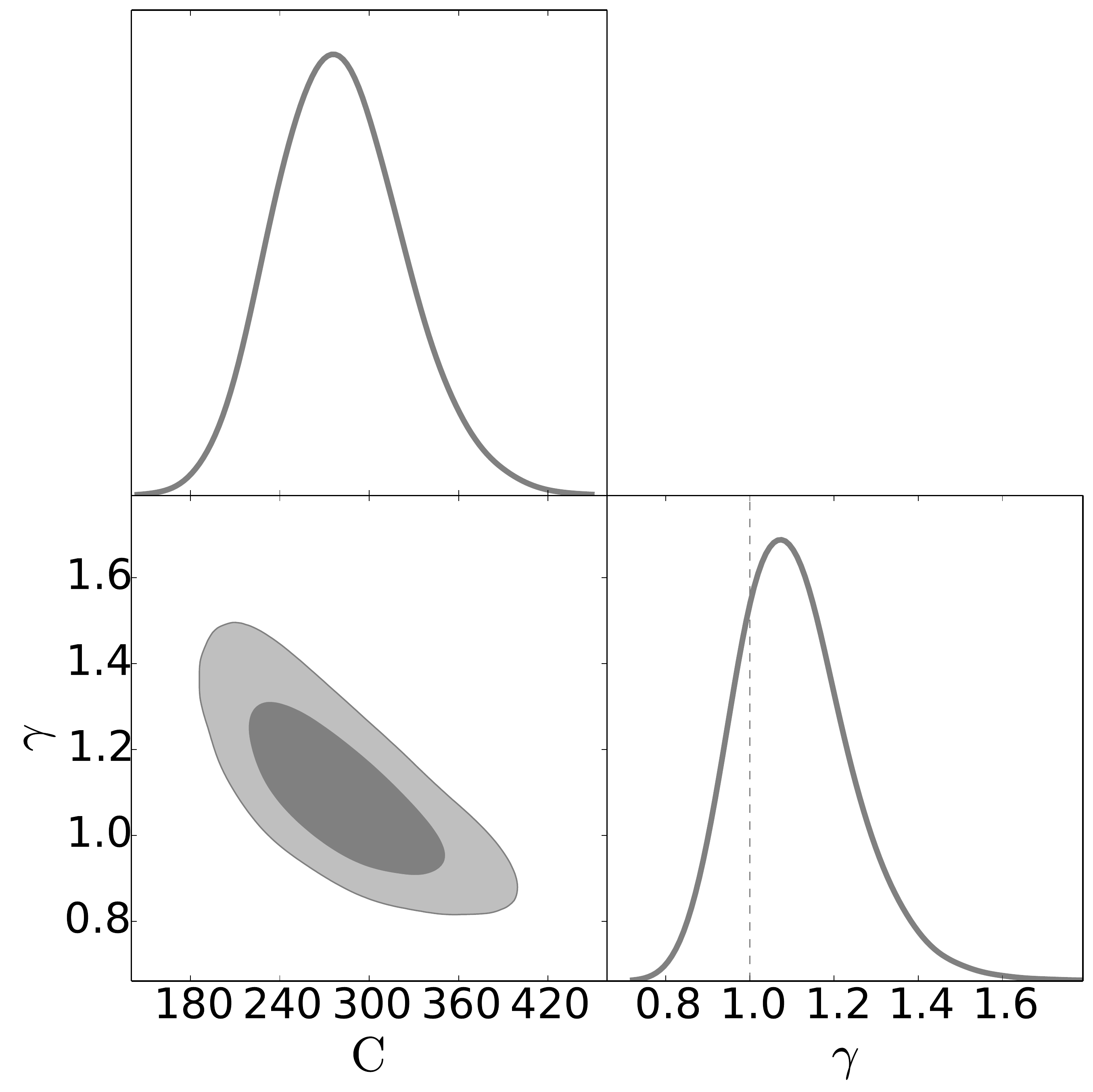}
\caption{Same as Fig.~\ref{fig:tridata} but for a simulated sample of 100 objects spanning $8<{\rm SNR}<\infty$.
\label{fig:trisim}}
\end{figure}

\begin{figure}[t!]
\centering
\includegraphics[width=\columnwidth]{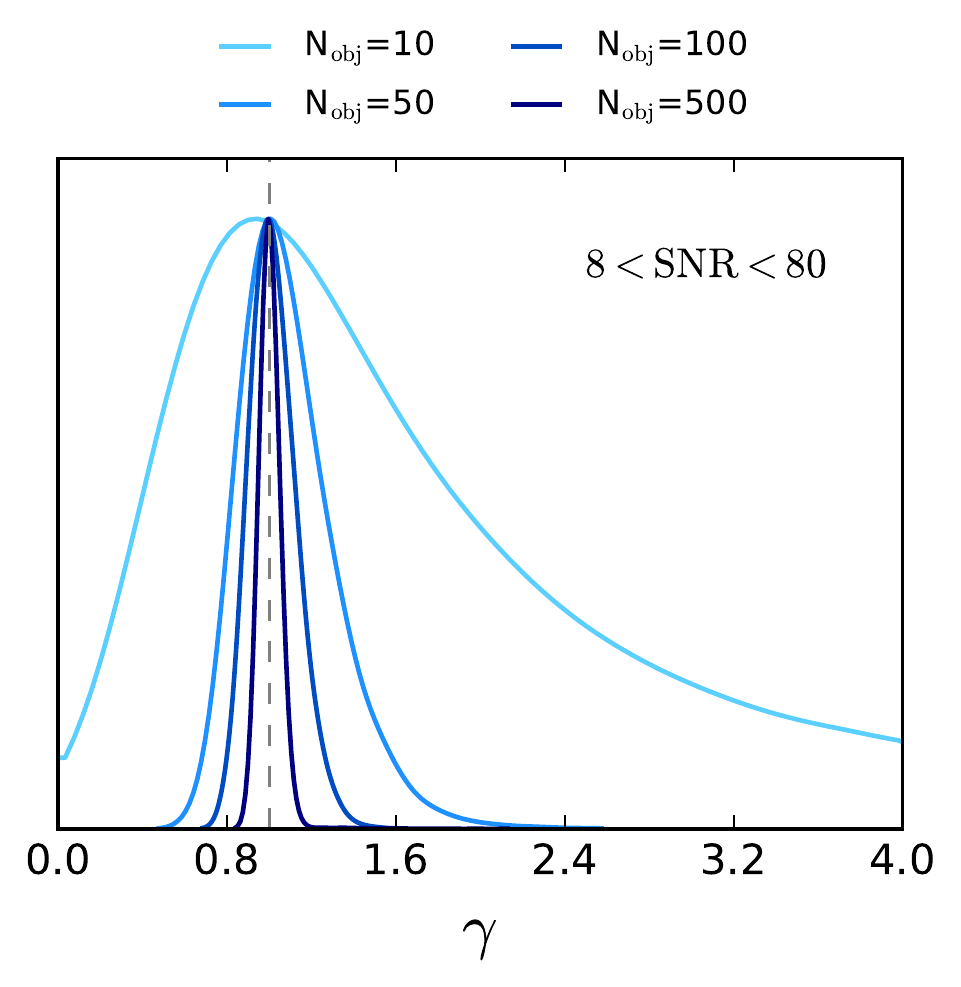}
\caption{Measurement of $\gamma$ as function of the number of objects detected. Curves from lighter to darker (distributions from wider to narrower) show the improvement achieved by increasing the number of sources.  
\label{fig:gamma}}
\end{figure}

It is important to notice that there are two main factors improving the $\gamma$ constraint: both the number of objets and the extended range in SNR contribute to the better measurement. We show this in Fig.~\ref{fig:gamma} where the impact on the constraint from increasing the number of objects in a wide, arbitrary chosen, range of SNR is summarized. This can be explained by thinking of a more numerous sample over a large $x$ range as a way to anchor the model and leave less freedom for a tilt. However, the model is going to 0 at high $x$ and detections with very high signal-to-noise contribute very little to the measurement, further additional information can be added by having more objects and better sampling at intermediate scales. 

To push the constraint even further and reach a percent level measurement of $\gamma$ will require $\sim$10000 objects, which, according to current event rate estimates, will be possible only extending the source sample to intermediate redshifts.

\section{Conclusion}
\label{sec:conclude}

Advanced LIGO detectors have opened a new window on the universe and provided a new probe of fundamental physics. After the first high-significance observation of GW from two black holes merging and other possible candidates during the first observing run in September 2015 of only 16 days, we expect hundreds of events per year once the design sensitivity has been reached. 

In this paper we have introduced a new method that uses the distribution of many gravitational wave sources (as we expect from future observations) in the nearby universe as a probe of the propagation of the gravitational wave signal. We test for the first time the way gravity propagates by bounding the power law scaling of the differential GW source counts, $dN/d\rho \propto \rho^{-3/\gamma-1}$, with $\gamma=1$ recovering the expected standard physics predictions. The model does not require any information on the astronomical counterpart to the gravitational wave events. With future insight into the source population distribution and evolution this method can also be extended to intermediate-to-high redshifts.

We first consider the case of few objects, 7 sources with a signal-to-noise from 8 to 24, as expected for lower-significance detections from Advanced LIGO. We find that, with only 7 gravitational wave sources in a narrow range of SNR, we can already impose a lower bound on $\gamma$, finding $\gamma>0.33$ at 95\% confidence level. We note that the distribution of our simulated sample is very consistent with the distribution of the trigger events observed by Advanced LIGO. We then simulate more observations to forecast the power of future data in constraining GW physics using this method. We find that we can achieve a sensitivity of $\sigma(\gamma)\sim 0.15$ with 100 sources over $8<{\rm SNR}<\infty$. We highlight how the constraint on $\gamma$ depends on the number of observed objects covering a wide SNR range.

This sensitivity will rule out, for example, relativistic MOND-like theories that would predict $\gamma\sim0.5$.

We have extended our derivation to include information on the chirp mass of the sources or on  the possibility of a gravitational wave signal decaying with a characteristic scale.  Once the Advanced LIGO team has provided the maximum detection distance as a function of black hole binary parameters, the formalism in this paper can also be used to constrain this length scale.

\section*{Acknowledgments}
EC and NB are supported by Lyman Spitzer Fellowships. We thank Neil Cornish, David Alonso and Vera Gluscevic for useful comments on the initial version of the paper posted, and Frans Pretorius for enlightening discussions.

\bibliography{ref}

\end{document}